\documentstyle[11pt,aaspp4]{article}

\slugcomment{Submitted to the Astrophysical Journal}

\lefthead{Trilling et al.}
\righthead{Orbital Evolution of Giant Planets}

\begin{document}

\title{Orbital Evolution and Migration\\
    of Giant Planets: Modeling Extrasolar Planets}

\author{D. E. Trilling}
\affil{Lunar and Planetary Laboratory, University of Arizona,
    Tucson, AZ 85721}

\author{W. Benz}
\affil{Steward Observatory,
	University of Arizona, Tucson, AZ 85721 and
	Physikalisches Institut, Universitaet Bern, Sidlerstrasse 5,
	CH-3012 Bern,
	Switzerland}

\author{T. Guillot}
\affil{University of Reading, Department of Meteorology,
	PO Box 243, Reading RG6 6BB, UK and
	Observatoire de la C\^ote d'Azur, BP 4229,
	06304 Nice Cedex 04, France}

\author{J. I. Lunine and W. B. Hubbard}
\affil{Lunar and Planetary Laboratory, University of Arizona,
    Tucson, AZ 85721}

\author{A. Burrows}
\affil{Department of Astronomy and Steward Observatory,
    University of Arizona, Tucson, AZ 85721}

\begin{abstract}

Giant planets in circumstellar 
disks can migrate inward from their initial (formation) 
positions.  Radial migration is caused by inward torques between
the planet and the disk; by outward 
torques between the planet and 
the spinning star; and by outward torques due to
Roche lobe overflow and consequent mass loss
from the planet.
We present self-consistent numerical
considerations of the problem of
migrating giant planets.
Summing torques on planets
for various physical parameters, we find 
that Jupiter-mass planets
can stably arrive and survive at small
heliocentric distances, thus reproducing
observed properties of some of the
recently discovered extra-solar
planets.   Inward 
migration timescales can be
approximately equal to or less
than disk lifetimes and star spindown
timescales. Therefore,
the range of fates of massive planets
is broad, and generally comprises three classes:
(I) planets which migrate inward too rapidly and lose all
their mass; (II) planets which migrate inward, lose
some but not all of their mass, and survive
in very small orbits; and (III) planets which do not
lose any mass. Some planets in Class III do not
migrate very far from their formation
locations.
Our results show that there is a wide range
of possible fates for Jupiter-mass planets
for both final heliocentric distance
and final mass.

\end{abstract}



\section{Introduction \label{introduction}}

\markcite{mb}
\markcite{bm}
\markcite{mq}
\markcite{butler97}
\markcite{cochran}
\markcite{noyes}
The recent discoveries
(Mayor \& Queloz 1995;
Marcy \& Butler 1996;
Butler \& Marcy 1996;
Butler et al. 1997;
Cochran et al. 1997;
Noyes et al. 1997)
of extra-solar planets
have revitalized discussions on the theory 
of planetary system formation and evolution.
In particular, several of these planets
are found to be on the order of a Jupiter-mass
($1\,M_{J}=2\times10^{30}$\,g) or greater,
and in very close proximity to their
central stars (Table~\ref{table1}). Previous to the 
discoveries of planetary companions in very
small orbits, it was predicted that
Jupiter-type planets would form
(and, by implication, exist)
only at or outside of the ice line
(3~-~5\,AU)\markcite{boss95}
(Boss 1995).
In addition, although close giant planet formation may be theoretically
possible\markcite{wuchterl}\markcite{wuchterl96}
(Wuchterl 1993, 1996), this requires the initial formation of a
solid core of at least 5 to 10 Earth masses which may be difficult 
to achieve very close to the parent star. It is therefore most likely 
that Jupiters cannot form at
small heliocentric distances\markcite{boss95}
\markcite{guillot}
(see also Boss 1995; Guillot et al. 1996),
which thus leaves the
question: How did the observed massive
close companions to several stars get to their
current locations?

\placetable{table1}

Jupiter-mass planets
can
migrate
inward from their formation
locations\markcite{lin96}\markcite{lp86}\markcite{lp93}
\markcite{gt80}\markcite{hourigan}
(see, for example, Goldreich \& Tremaine 1980;
Ward \& Hourigan 1989; Lin \& Papaloizou 1986;
Lin \& Papaloizou 1993;
Lin, Bodenheimer, \& Richardson
1996).  We consider under what physical
conditions planets may migrate inward,
and how, where, why, and when a planet will
stop its inward migration.
Rather
than consider catastrophic interactions
among more than one planet\markcite{weid}\markcite{rasioford}
(Weidenschilling \& Marzari 1996;
Rasio \& Ford 1996),
we model isolated massive planets which smoothly migrate
as a result of the net torques
on the body.
We find three
broad classes for massive
planets: (I) planets which migrate inward
very quickly and disappear due to mass loss from
Roche lobe overflow; (II)
planets which migrate inward, lose
some but not all of their mass, and stably
survive at small heliocentric distances;
and (III) planets which do not lose
any mass during migration. Some planets
in Class III move radially only 
a small distance from their formation
locations. In the context 
of this picture, massive
close companions can result from a variety
of initial conditions, and form a population
which overlaps with and includes the detected
close companion
population.  Our distribution of final masses
and heliocentric distances predicts that massive
planets can be present at any heliocentric distance
between their formation locations and extremely
small orbits, and we predict that as detection
sensitivities increase, massive planets will be found
to have an almost continuous distribution of
heliocentric distances, from a fraction of an
AU all the way out to the ice line or beyond.

Our general model is to sum up the torques on
a planet and find its radial motion in the
circumstellar disk.  We compute,
in a one dimensional model, torques on
a planet due to disk interactions (inward
for the parameters which we adopt);
torques on a planet due to interactions with
the rapidly-spinning star (outward); and 
torques on the planet due to mass loss onto
the star (outward for the conservative
mass transfer case).  Sections \ref{disk},
\ref{tides}, and \ref{massloss},
respectively, 
describe the calculation of
these torques and the resulting
radial motions.
These torques must be calculated 
numerically in order to combine the 
results in a self-consistent
model. In Section \ref{results},
we present the results of adding these
torques together and calculating the orbital
evolution of massive planets.
Our work is the first to quantitatively
and self-consistently consider these three
torques and the problem of extrasolar planets.
Lastly, in Section
\ref{discussion} we discuss our results.
Section \ref{conclusions} provides
conclusions
and predictions.

\section{Planet Formation and Assumptions in the Model \label{assumptions}}

Our model begins, at time = 0, with a fully formed
gaseous giant planet of a given mass at a given heliocentric distance.
The circumstellar disk is assumed to have smooth,
power-law radial density and temperature profiles,
and to be axisymmetric.
Jupiter-mass planets may require
most of the lifetime of the disk to accrete
($10^{6}$ to $10^{7}$\,years)\markcite{zuck}\markcite{pollack}
(Pollack et al. 1996; Zuckerman, Forveille,
\& Kastner 1995).
However, we incorporate this possibility in our model
by
allowing some disks
to dissipate during the planet's migration,
thus letting the
effective disk lifetime ($\tau_{disk}$)
be the actual disk lifetime
minus the time to form the planet,
and consider this quantity to be the
limiting time constraint.

When an accreting planet has 
sufficient mass, it will form
a gap in its disk; gap formation
terminates the accretion process.
In all interesting
cases in this study, the gap forms quickly.
Therefore, starting with a fully formed
giant planet in a disk with an
initial smooth density
distribution is adequate and appropriate. 
We solve for the one-dimensional
(radial) movement of massive
planets.
The orbits are assumed
to be Keplerian and circular 
at all times.

\section{Disk Evolution and Disk Torques \label{disk}}

We model a thin axisymmetric disk
with radial temperature
and initial surface density
profiles given by power-laws with
exponents -1/2 and -3/2, respectively\markcite{takeuchi}
(Takeuchi, Miyama, \& Lin, 1996).
Our nominal circumstellar disk
has the following physical parameters:
$\alpha$ equal to $5\times10^{-3}$, where
$\alpha$
is the usual disk viscosity parameter 
defined by
$\alpha = \nu\Omega/c_{s}^{2}$,
in which $\nu$ is the 
viscosity in the disk,
$\Omega$ is the Keplerian
angular velocity at a given heliocentric
distance, and $c_{s}$ is the sound
speed\markcite{shakura}
(Shakura \& Sunyaev 1973);
$M_{disk}=1.1\times10^{-2}\,M_{\sun}$\markcite{bs}
(Beckwith \& Sargent 1993);
and scale height equal to 0.05\,AU
at 5.2\,AU\markcite{takeuchi}
(Takeuchi et al. 1996).
We take the nominal effective lifetime of the
disk to be $10^{7}$\,years, based
on observations (Zuckerman et al. 1995),
with a range of $10^{6}$ to $10^{7}$\,years.
All of 
these physical parameters are varied
from their nominal values to model 
disks with different properties.

A massive planet and the circumstellar
disk
interact tidally which results in
angular momentum transfer
between the disk and the
planet\markcite{gt80}\markcite{hourigan}
\markcite{lp93}\markcite{takeuchi}\markcite{lp86}\markcite{ward}
(e.g., Goldreich \& Tremaine 1980; Lin \& Papaloizou 1986, 1993;
Ward \& Hourigan 1989;
Takeuchi et al. 1996;
Ward 1997a; Ward 1997b). The
planet's motion in the disk
excites density waves both
interior and exterior to
the planet.
These waves create a gap in
the disk as the 
planet clears material from
its orbit. The size of the
gap depends on the viscosity of the 
disk and inversely on the mass of the 
planet\markcite{takeuchi}\markcite{lp86}
(Lin \& Papaloizou 1986; Takeuchi et al. 1996;
see eqs. \ref{dadt_disk} - \ref{sigma}, below).
The magnitude of the torque depends
on the amount of disk material present
near the planet,
and thus on the size of the gap.
Torques on the planet are
caused by gravitational interactions
between the planet and density waves
which occupy Lindblad resonances
in the disk. We simplify the problem 
by adopting the 
impulse approximation of\markcite{lp86}
Lin \& Papaloizou (1986), 
in which dissipation of the
density waves is
assumed to be local and
angular momentum
is deposited close to the
planet through dissipative
phenomenon such as shocks.
The local dissipation approximation 
is reasonable because the disk
which is nearest the planet
has the largest influence in causing
a torque on the planet. The accuracy
and validity of this approximation
is discussed at the end of this section.

The radial motion of a planet
in a circumstellar disk, due
to interactions with the disk,
is given by

\begin{equation}
\frac{da}{dt}=
-\left(\frac{a}{GM_{\star}}\right)^{\frac{1}{2}}
\left(\frac{4\pi}{M_{p}}\right)
\int^{R_{out}}_{R_{in}} R \Lambda\left(R\right) \Sigma\left(R\right) dR
\label{dadt_disk}
\end{equation}

\noindent where $a$ is the heliocentric distance
of the planet;
$M_\star$ and $M_p$
are the masses of the central star and the 
planet, respectively; $R$ is the radial coordinate,
with $R_{in}$ and $R_{out}$
are the inner and outer boundaries
of the disk, respectively; $\Lambda$ is the 
injection rate of angular momentum
per unit mass into the disk
due to interactions between
the disk and the planet;
and $\Sigma$ is
the surface density 
of the disk\markcite{lp86}
(Lin \& Papaloizou 1986).
Note that the 
radial motion of the planet
is inversely proportional to the
mass of the planet, so that more massive
planets move less rapidly.
The angular momentum injection rate
for the impulse approximation
with local dissipation is also taken
from\markcite{lp86}
Lin \& Papaloizou (1986):

\begin{equation}
\Lambda\left(R\right)=
{\rm sign} \left(R-a\right)
\frac{fq^{2}GM_{\star}}{2R}
\left(\frac{R}{|\Delta_{p}|}\right)^{4}
\label{lambda}
\end{equation}

\noindent where $f$ is a constant of
order unity, $q$ is the mass ratio
between the planet and the star
($M_{p}/M_{\star}$),
and $\Delta_{p}$ is equal to the
greater of $H$ or
$|R - a|$, where $H$ is
the scale height of the disk.
We solve for 
$\Sigma$ in equation~\ref{dadt_disk} with
a fully implicit solution to the
continuity equation
for the disk rewritten as\markcite{lp86}
(Lin \& Papaloizou 1986)

\begin{equation}
\frac{\partial \Sigma}{\partial t} =
\frac{1}{R}\frac{\partial}{\partial R}
\left[3R^{1/2}\frac{\partial}{\partial R}
\left(\nu\Sigma R^{1/2}\right) -
\frac{2\Lambda\Sigma R^{3/2}}{\left(GM_{\star}\right)^{1/2}}\right].
\label{sigma}
\end{equation}

\noindent We take the density equal to zero
in the innermost zone of the disk to represent 
material from the inner boundary of the disk
falling onto the star.
Thus, we solve for $\Sigma$(R)
and
$\Lambda$(R) and then compute
the integral in
equation~\ref{dadt_disk}
to find the radial motion
of the planet.

The gap formed by the planet
in the disk is crucial
in determining the 
behavior of the system.
Figure~\ref{gap} shows gap size in 
the disk as a function of initial
planetary mass. The gap is defined
as the region in which the surface
density is less than half of what
it would be if there were no planet
in the disk. This figure
compares our results (circles), using the 
local dissipation approximation
of Lin \& Papaloizou (1986), against
results using the more computationally intensive
WKB approximation of Takeuchi et al. (1996)
(shown as squares).
For each model, $\alpha$'s of
$5\times10^{-3}$ and $10^{-2}$
are shown. The impulse approximation
which we use closely reproduces 
gap formations found with the more 
complex scheme; therefore our 
simplifying assumption is valid.
The time ``snapshot'' shown in
Figure~\ref{gap} is $10^4$\,years,
a time after the gap has fully formed
but before the planet has started
to migrate significantly from its initial
location. The disk continues to
evolve after this time, and as the 
planet moves inward, the gap created
by the planet moves inward as well.
The inner edge of the gap is continually
eroded, and the outer edge fills in
viscously, as the planet moves inward.
Note that larger viscosities cause
smaller gaps and therefore faster inward
migration; larger diskmasses also cause
smaller gaps and subsequently more rapid
inward migration.

\placefigure{gap}

We assume that the circumstellar disk dissipates
after $10^7$\,years, based on observational evidence
(Zuckerman et al. 1995). In our model, we do not 
attempt to model the exact physics of disk dissipation.
Instead, we simply assume that, as a local phenomenon,
the disk has disappeared from regions close to the 
planet after this time. In terms of the behavior of 
a planet, only the local disk is important, and disk-clearing
processes elsewhere do not affect the
behavior of the planet. However, in future work,
we intend to model disk dissipation explicitly.

\section{Torques from the Spinning Star \label{tides}}

As a migrating planet gets close to its central star,
tidal bulges raised on the star by 
the planet become important to
the net radial motion of the planet.
Because stars are dissipative,
the stellar tidal bulge is not aligned with the line
of the centers of mass of the star and the
planet\markcite{gs}\markcite{stacey}\markcite{hubbard}
(Goldreich \& Soter 1966;
Stacey 1977; Hubbard 1984). 
In most cases, our model planets do not migrate
to separations smaller than the
co-rotation point with the rapidly rotating star;
therefore,
we assume that in all cases,
the star is rotating faster than the
orbital period of the planet, so that
the tidal bulge leads the line of centers.
The torque on the planet in this case is in the outward
sense, and the planet slows the rotation rate of the star,
as is the case with the Earth-Moon system.
As energy is dissipated within the star and the
star slows down, the planet
must move outward to conserve angular momentum.

The radial motion of the planet due to spin torque 
interaction with the star is given by

\begin{equation}
\frac{da}{dt}=
\frac{9}{2}
\Omega_{p} q
\left(\frac{R_{\star}}{a}\right)^{4}
\frac{R_{\star}}{Q_{\star}}
\label{dadt_star}
\end{equation}

\noindent where 
$\Omega_{p}$ is the Keplerian angular velocity
of the planet, $R_{\star}$ is the stellar radius, and
$Q_{\star}$ is the tidal dissipation factor
of the star\markcite{lin96}\markcite{gs}
(Goldreich \& Soter 1966; Lin et al. 1996).
The numerical coefficient is from the
expression for body of uniform density;
however, departures from this can be incorporated into
uncertainties in $Q_{\star}$ (see below).
Including the effects of infall of material onto the star and
the release of energy from gravitational
contraction, the stellar radius
decreases from early times until $10^{7}$\,years with
the following function:

\begin{equation}
R_{\star}\left(t\right)
=\left[\left( \frac{28 \pi \sigma T^{4} t}
{GM_{\star}^{2}} \right) +
\frac{1}{R_{o}^{3}} \right]^{-\frac{1}{3}}
\label{rstar}
\end{equation}

\noindent where
$\sigma$ is the Stefan-Boltzmann constant, $T$ is the surface 
temperature of the star
(taken to be 4600\,K), $t$ is elapsed time since 
the star completed its accretion, and $R_{o}$ is the
star's initial radius,
taken to be four times the sun's current
radius\markcite{cameron}
(Cameron 1995).

The dissipation factor $Q$
describes how efficiently
rotational energy is dissipated by friction
within an object\markcite{hubbard}\markcite{rasio}
(Hubbard 1984; Rasio et al. 1996).  We use
a value of $1.5\times10^{5}$ for
$Q_{\star}$\markcite{lin96}
(Lin et al. 1996),
and a range of $Q_{\star}=1.5\times10^{4}$
to $1.5\times10^{6}$, for a
pre-main sequence star.  By comparison, Jupiter has
$Q_{J}$ in the range $10^{4}$ to $10^{5}$\markcite{hubbard}
(Hubbard 1984), and a main sequence star has
$Q_{\star} \geq 10^{6}$. For smaller $Q_{\star}$'s, the star dissipates
energy less efficiently, and the (outward) tidal torque
on the planet is greater than is the case for larger
$Q_{\star}$'s.

We adopt a spin-down time for the star of $10^{8}$\,years,
since after this time, the star's
rotation rate
has decreased by roughly an order of magnitude
from its rapidly rotating state\markcite{sku}
(Skumanich 1972).
The effect of different spindown timescales,
from $10^7$ to $10^8$\,years,
on the results of the model is small.
The tidal torque is essentially off
after $10^8$\,years due to
the star's
higher $Q_{\star}$ and slower rotation
rate.
In fact, all systems with close companions
(with the exception
of the tidally locked $\tau$ Boo system)
are still evolving dynamically due to tidal torques
between the bodies,
but the orbital decay time for 51~Peg~b,
for example, is more than $10^{12}$\,years\markcite{rasio}
(Rasio et al. 1996),
much longer than the main sequence
lifetime of the star,
so we do not include this late orbital
evolution in our model.

Tidal heating of the planet is caused by
tidal bulges raised on the planet due to the 
star, the same mechanism which produces outward
tidal torques on the planet.
Significant heating of a planet due to external torques would alter
its internal structure and behavior
and therefore be important in the 
mass overflow regime. The tidal heating rate 
of the planet is
given by\markcite{tittemore}
Lunine \& Tittemore (1993) as 

\begin{equation}
\frac{dE}{dt}
= - \frac{k_{2}M_{\star}n^{3}R_{p}^{5}}{a^{3}Q_{p}}
\left( \frac{21}{2}e^{2} + \frac{3}{2}\theta^{2} \right)
\label{dedt}
\end{equation}

\noindent where $E$ is the energy input
to the planet 
due to tidal heating,
$k_{2}$ is the planet's Love number
($k_{2} \sim$ 0.5\markcite{hubbard} 
(Hubbard 1984)), $n$ is the orbital mean motion, $e$ is
the eccentricity of the orbit, and $\theta$
is the obliquity.
For a planet with a four-day period
and $e$ = 0.1 and $\theta$ = 0,
the heat energy input is $\sim 10^{26}$\,ergs/sec.
Since
a Jupiter-mass planet in such
a close orbit would have 
luminosity $\geq 10^{28}$\,ergs/sec
(see Figure~\ref{hr}),
we find that tidal heating of
a close planetary companion
is negligible.

\section{Mass Loss and Conservation of Angular Momentum \label{massloss}}

When a migrating planet gets sufficiently
close to its primary star, the planet's
radius can exceed its Roche radius. When
this occurs, mass transfer from the planet
to the star takes place. During transfer,
the planet moves outward to conserve
the angular momentum of the system.
In the case of stable mass transfer,
the planet moves
to a distance at which its planetary 
radius is equal to the 
Roche radius. Therefore,
the distance
to which the planet will move is determined
by the planetary radius, and hence is 
a function of the intrinsic
properties of the planet: its
age, temperature, and mass.

The planets in
our model are not point masses, but
have radii and internal structures that
are calculated at each location and time
through the planet's evolution.

\subsection{Internal Structure of Model Planets \label{internal}}

A grid of quasi-static evolution models
was calculated as described in\markcite{guillot}
Guillot et al. (1996). For a given composition, the radius
of a
planet ($R_{p}$) is a function of its mass $M_{p}$,
its equilibrium temperature
$T_{eq}$ and time $t$. The grid was calculated
for 0.4\,$M_{J}\le
M_{p}\le 10\,M_{J}$, $0\,K\le T_{eq}\le 2000$\,K, and
for a few billion years, starting from an
initial extended planet of about 16 Jupiter radii
($R_{J}$)
at $t=0$.

When the incoming stellar heat is redistributed equally over the entire
atmosphere of the planet, the equilibrium temperature is related to
the orbital distance by:

\begin{equation}
4\pi\sigma T_{\rm eq}^4=(1-A)\frac{L_\star}{4a^2}
\label{teq}
\end{equation}

\noindent where $A$ is the Bond
albedo of the planet and $L_\star$ is the stellar luminosity. For
simplicity, we assume $A=0$ and $L_\star=L_{\sun}$
($1\,L_{\sun}=3.9\times10^{33}$\,erg/sec).
Of course, any uncertainty on $A$ is equivalent
to an uncertainty on $L_\star$ (except
in the limit $A\rightarrow 1$). The
effect of those is discussed below. 

The effective temperature of the planet is defined by the equation

\begin{equation}
4\pi R_{p}^2\sigma T_{eff}^4=L_{p}+4\pi R_{p}^2\sigma
T_{eq}^4
\label{teff}
\end{equation}

\noindent where
$L_{p}$ is the intrinsic luminosity of the planet, and is a
function of $M_{p}$, $T_{eq}$ and $t$. The time variable is
inappropriate as a conserved quantity during 
mass loss calculations because the interior of the
planet and the orbital evolution are not coupled,
and because a time origin cannot be defined in an
absolute way. Heliocentric distance is not a useful
conserved quantity either, since $a$ changes 
while the planet moves. Therefore, although the radius of the
planet is, in general, $R_{p}(M_{p},a,t)$,
when considering continuous mass loss from a planet,
we use $R_{p}(M_{p},T,S)$, where
$S$ is the specific entropy (entropy per unit mass).
Planetary mass, temperature, and entropy are
continuous variables over which the grid of
models can be both interpolated and differentiated;
temperature and entropy are both conserved 
during mass loss events.
The grid of models has minimum mass of 0.4\,$M_{J}$; 
therefore, planets are considered to have lost
all their mass and disappeared when $M_{p}<0.4\,M_{J}$.
Note that none of our planets therefore ever 
enter a regime in which the planetary core is involved
in mass transfer. The core would be made of solid (rocky)
material, and behave very differently in a mass transfer
regime; we intend to study this behavior in future
work.

The conservation of the
specific entropy at the center of the planet during mass-loss
events is exact in the case of a
fully convective planet, as long as the atmosphere can adjust to the
new equilibrium on a time scale much shorter than the mass loss time
scale. This is generally the case since for $T\approx 1000$\,K
and pressure $\approx 1$\,bar the heat capacity is $c_p\approx
10^8$\,erg/K\,g and the Rosseland mean opacity is
$\kappa_{\rm R}\approx 10^{-2}\rm\,cm^2/g$, so that the
radiative diffusivity can be estimated as:

\[
K_\nu=\frac{1}{\rho c_p}\frac{16\sigma T^3}{3\rho\kappa_{\rm
R}}\approx 10^9\rm\,cm^2/s.
\label{kappa}
\]

\noindent Therefore, the corresponding time scale is $\tau_\nu\approx
L^2/K_\nu\approx 0.3$\,years, using $L\approx 10^8$\,cm (about 1\%
of the planetary radius). This is smaller than the mass loss
time scale since it takes about $10^3$ to $10^4$\,years to lose about
1\% of the planetary mass by Roche lobe overflow (see below).

However, planets in very close orbit do not generally stay fully
convective but develop an inner radiative zone due to a strong
(and unavoidable) decrease of their
internal luminosity\markcite{guillot}\markcite{guillotproc}
(see Guillot
et al. 1996, 1997). The entropy is therefore larger at the top than at
the bottom of the radiative zone. As a result, not only the
atmosphere but also
the inner radiative zone have to adjust to the new equilibrium. The
previous time scale estimation, now estimated using twice the
temperature and a 
mean pressure of 100\,bar, yields a $\tau_\nu$
which is $\sim$300 times larger
(assuming the same characteristic opacity),
or about 100\,years. This is close to the mass
loss characteristic time scale: non-equilibrium effects may
therefore be
significant. We will neglect them in the present work, thereby
somewhat underestimating the extent of mass loss by Roche lobe
overflow. Other sources of uncertainty are expected to be more
significant, as discussed hereafter.

The largest source of uncertainty in the calculation of the internal
structure and evolution of the planets and therefore on the presence
and magnitude of mass loss torques is due to the inaccurate
representation of the atmospheres of these objects. In our
calculations, mass loss always occurs when the planet is very close to
its parent star. The incoming stellar heat flux is then very
significant, and the intrinsic luminosity of the planet comparatively
very small. Interior and atmospheric models
are coupled
using a relation linking the effective temperature to the 
temperature at a given pressure level (see e.g.\markcite{saumon}
Saumon et
al. 1996). This relation is based on atmospheric models calculated
assuming {\it no} incoming stellar flux, and is equivalent to
assuming that the stellar flux (mostly emitted at short wavelength) is
absorbed deep in the atmosphere. Although it is not totally
unrealistic (and it is the best that can be done at the moment), the
resulting $R_{p}(M_{p},a,t)$ relation is very uncertain
(i.e., the accuracy on the partial derivative $dR/dt$
is probably only about
a factor two), thus affecting when mass loss occurs and how much mass is
effectively lost by the planet. It is also
important to emphasize that the radii are defined at the 10\,bar
level, and that ``hot Jupiters'' might have extended atmospheres, so
that the effective radius of the planet is slightly larger than
estimated here, although the amount of mass contained in
the extreme upper atmosphere is small. 

Finally, the uncertainty in $A$ and $L_\star$ will tend
to shift the orbital distances at which mass loss occurs.
For example, a stellar luminosity of
$2L_{\sun}$ causes a planet to lose mass at a
larger heliocentric distance than for the nominal case.
The result is that in the $2L_{\sun}$ case, the planet's 
final mass is smaller than for the same initial
mass planet in the nominal disk.
However, we stress that
refinements in
the treatment of planetary atmospheres will not 
affect the qualitative behavior of the system,
although the exact values of mass and distance at
which mass loss occurs may change with improvements
in planetary atmosphere models.

\subsection{Planetary Mass Loss and Consequent Radial Motion}

The Roche lobe overflow regime 
begins at the distance where the planet's radius is
equal to or greater than its Roche lobe radius.
The Roche radius is given by
the following\markcite{egg}
(Eggleton 1983):

\begin{equation}
R_{L}=
\frac{0.49q^{2/3}a}
{0.6q^{2/3} + \ln \left(1 + q^{1/3}\right)}.
\label{roche}
\end{equation}

For stable mass transfer in which
material from the planet is transferred inward
onto the central star,
the mass loss rate due to Roche lobe
overflow of a planet is 
determined by and balances the net inward torque,
and is given by\markcite{ci}\markcite{benz}

\begin{equation}
\frac{dM_{p}}{dt}=
C^{-1}M_{p}\left(\frac{d\ln J_{p}}{dt}-
\frac{1}{2}\frac{d\ln R_{p}}{dt}\right)
\label{massloss1}
\end{equation}

\noindent where $J_p$ is the orbital angular momentum 
of the planet, and where

\begin{equation}
C=
\left(1 - q\right)-\frac{1}{2}\beta\left(1+q\right)+
\frac{1}{2}\alpha_{p}
\label{massloss2}
\end{equation}

\noindent (Cameron \& Iben 1986;
Benz et al. 1990). The derivative
of the planetary radius with
respect to time in equation~\ref{massloss1} is 
at constant mass, but allows
heliocentric distance and
age to change. Since 
temperature is a function of distance
and entropy
is the time variable, we have

\begin{equation}
\frac{d \ln R_{p}}{dt} =
\frac{\partial \ln R_{p}}{\partial S}
\frac{dS}{dt} + 
\frac{\partial \ln R_{p}}{\partial T}
\frac{dT}{da}
\frac{da}{dt}.
\label{partials}
\end{equation}

\noindent The planetary radius partial
differentials come from our grid of
atmosphere models; $dS/dt$ comes from
the evolution models as well; $dT/da$
comes from equation~\ref{teq};
and $da/dt$ is the planet's radial
motion from the sum of disk and tidal torques.
The remaining parameters in 
equation~\ref{massloss2} are

\begin{equation}
\beta = \left(\frac{d \ln R_{L}}{d \ln q}\right)_{a}
\label{beta}
\end{equation}

\noindent for $R_{L}$
from equation~\ref{roche};
and the parameter $\alpha_{p}$, which is the
exponent in the mass-radius relation,
and should not be confused with the
disk viscosity $\alpha$. The mass-radius
relation exponent $\alpha_{p}$,
which is the derivative of planetary
radius with respect to planetary
mass at constant temperature
and entropy, is
given
by

\begin{equation}
\alpha_{p} =
\left(\frac{d \ln R_{p}}{d \ln M_{p}}\right)_{T,S}.
\label{alpha}
\end{equation}

\noindent We calculate $\alpha_{p}$ from our grid of
atmosphere evolution and structure models.
Note that $\alpha_{p}$ in this work
is defined differently than in\markcite{ci}\markcite{benz}
Cameron \& Iben (1986) and Benz et al. (1990),
by a minus sign.

In the mass-age-temperature regime
in which we consider mass loss (planets
with $0.5\,M_{J}<M_{p}<10\,M_{J}$,
temperature~$>$~1000\,K, and age
less than $10^{8}$\,years),
a planet instantaneously
losing mass
at
constant T and S
expands.
In order to conserve 
the angular momentum
of the system, a planet
which loses mass to its primary must move outward.
A planet losing mass is pushed
outward while $R_{p} > R_{L}$,
and the 
planet will lose mass until

\begin{equation}
R_{L}^{\prime} = R_{p}^{\prime} \hspace{3ex}
{\rm for} \hspace{3ex}
M_{p}^{\prime} < M_{p} \hspace{3ex}, \hspace{3ex}
a^{\prime} > a \hspace{3ex} , \hspace{3ex}
{\rm and} \hspace{3ex}
R_{p}^{\prime} > R_{p}
\label{prime}
\end{equation}

\noindent where the primes refer to
values at the planet's new mass and
heliocentric distance.
Since the Roche radius is
proportional to heliocentric distance
(eq.~\ref{roche}),
any subsequent
inward motion of the planet
decreases the Roche radius
further,
which results in more mass loss from the planet.
Since the circumstellar
disk provides a torque which
pushes the planet inward,
as long as the disk is
present, mass loss proceeds 
continuously at the planetary Roche radius,
once it has started.
The planet
will continue to lose mass and move to larger and
larger heliocentric distances 
to conserve angular momentum. 
The timescale for mass loss from
the planet onto the star can be quite short,
a few times $10^{6}$ years or less.
However, when the disk dissipates,
the inward torque goes to zero, and 
the mass loss stops. If a planet
is in the process of losing mass
when the disk dissipates, mass
loss will cease, and the planet will be 
stranded at a distance
where $R_{L}$ = $R_{p}$,
with $M_{p}<M_{i}$.
The model stops during the mass
loss regime when the planet's mass
becomes less than 0.4 $M_J$, or when the 
model stops converging during mass loss,
typically as the mass approaches 0.4 $M_J$.

\section{Results \label{results}}

The general approach involves studying the evolution
of a planet with time and distance. Specifically,
we
sum up the torques on a planet
at a given time and distance, and solve for
the planet's motion from this sum of torques.
The torques are
due to interactions with the disk (inward); due to the spin of 
the star (outward); and due to mass loss (outward).
Additionally, at every time and distance, we solve
for the planet's radius.

There are several major stages in a planet's migration.
The first stage, which lasts from formation
time (time zero) for $\gtrsim 10^{6}$\,years,
is the inward migration stage. The only important
torques on the planet arise from disk interactions.
The second stage occurs if or when a planet 
reaches a small enough heliocentric distance
for spin (tidal) torques to become important,
in addition to disk torques. This second stage
can last for $10^{6}$ to $10^{7}$\,years.
In this stage, inward torques from the
disk are somewhat offset by outward torques
from interactions with the spinning star.
A third stage, which not all planets reach,
begins when the planet's radius exceeds the
Roche radius, and the planet starts to lose 
mass. The onset of mass loss
replaces inward motion with outward
motion. This third stage is short
because mass loss is rapid, but if it occurs
close to t~=~$10^{7}$\,years, it is possible
that the disk may dissipate while the planet
is losing mass, thus leaving a planetary remnant
at a small heliocentric distance. Stage four
corresponds to times between $10^{7}$ and
$10^{8}$\,years, when a planet
may be subject to outward torques due to star (spin)
tides, but neither disk nor mass loss torques 
are present. Planets can therefore only
move outward during this stage.
Finally, the last stage is
after $10^{8}$\,years, after which spin torques also
turn off, and only very small late stage migration
may occur. This late stage migration is also 
due to tidal interactions between the star 
and the planet, but for most planets, occurs
on a timescale longer than the age of a
main-sequence star\markcite{rasioetal}
(Rasio et al. 1996), and is negligible. We do
not consider this late-stage tidal migration
in our model.

There exist three broad classes
of fates for migrating planets.
In Class I, planets migrate
inward very rapidly and lose all their
mass onto the star by Roche
lobe overflow. Planets in Class II
migrate to small heliocentric distances and
lose some, but not all, of their mass onto
the central star. These planets are losing
mass when the disk dissipates, terminating
mass transfer from the planet. These
planets survive until $10^{10}$\,years
at small heliocentric distances and 
masses smaller than their initial masses.
Planets in Class III do not lose any
mass during their evolution.
The more massive of these planets do not migrate
very far, and reside at distances
very close to their formation
locations.

The nominal case for our model has the following
values for the variable parameters:
$\alpha=5\times10^{-3}$\markcite{takeuchi}
(Takeuchi et al. 1996);
diskmass~= $1.1\times10^{-2}\,M_{\sun}$\markcite{bs}
(Beckwith \&
Sargent 1993);
$\tau_{disk}~=~10^{7}$\,years\markcite{zuck}
(Zuckerman et al. 1995);
$\tau_{spin}~=~10^{8}$\,years\markcite{sku}
(Skumanich 1972);
and $Q_{\star}=1.5\times10^{5}$\markcite{lin96}
(Lin et al. 1996).
All of these parameters were varied in turn
to determine which physical parameters are most
important to the evolution and migration of a planet.

Table \ref{nominalmass} shows the results
of planets with different initial masses in
the nominal disk. The boundaries between
the different classes are shown quite clearly,
as planets with initial masses $\leq3.36\,M_{J}$
are Class I planets, which lose all their
mass; planets with initial masses
$3.36\,M_{J}<M_{p}<3.41\,M_{J}$
are Class II planets which lose some of their
mass but survive to $10^{10}$\,years;
and planets with initial masses $\geq3.41\,M_{J}$
are Class III planets which do not lose any mass.

\placetable{nominalmass}

Figures \ref{at} - \ref{ma}
correspond to Table~\ref{nominalmass}
and show the results of models for the 
nominal case, with various initial masses.
Figure~\ref{at} shows heliocentric distance versus
time, and in particular shows that migration timescales
are a few times $10^{6}$ years. For planets 
in Class III, the dominant effect is that
the planets clear a large gap in the disk
on a timescale shorter than the migration
timescale,
so that the inward disk torque is very 
small, and the planets do not move very
far.
In general, larger planets
are generally not found at small
heliocentric distances, although
deviations from the nominal disk do
allow for this to happen.
Planets' inward motions are
slowed by the outward tidal torque,
seen as the region at small heliocentric
distances with significantly shallower,
but still negative,
slopes. Planets' inward motions
are reversed if the
planetary radius exceeds the Roche radius
and mass loss begins.
Figure~\ref{mt} shows mass versus time, 
and in particular, clearly distinguishes 
planets in Class I, which lose all their
mass, as well as the members of Class II which
lose some, but not all, of their mass.
Figure~\ref{ma}
shows the masses of the migrating planets 
versus heliocentric distances. As described 
above, as planets lose mass, they move
away from their central stars,
seen as a line with a negative slope in
this figure. The sign of this slope 
is the same as the sign of $\alpha_{p}$.

\placefigure{at}
\placefigure{mt}
\placefigure{ma}

Figure~\ref{mfmi} shows final mass versus 
initial mass for planets in the nominal
circumstellar disk.
The
three classes of planets
are separated by the vertical
lines in the figure.
As seen in this figure, there
is a small range of initial masses
for which a planet has a finite final
mass less than its initial mass
(Class II).
We call the range of initial masses,
for a given set of disk
parameters, for which
$0<M_{f}/M_{i}<1,$
the critical mass range.
For the nominal disk, this
critical mass range 
spans initial masses from
3.37 to 3.40\,$M_{J}$,
inclusive. 
In all cases,
the critical mass range would be
broader if the mass loss
rate from the planets were slower.
A smaller mass loss rate would
result from smaller net
inward torques, reducing the 
$d \ln J_{p}/dt$ term; or
from a planet's having a
modified internal structure, thus
changing the $d \ln R_{p}/dt$ term
(in eq.~\ref{massloss1}). See the 
discussion section for further discussion
on the width of the critical mass range.

\placefigure{mfmi}

Some insight is gained
by examining the evolution of
migrating planets on 
a Hertzsprung-Russell diagram
(Figure~\ref{hr}).
In such a diagram,
one can easily see the stages of evolution
of a migrating planet. At early times,
the evolution of a migrating massive planet
is dominated by contraction of the giant planet,
during which time the evolution track follows
the boundary of the Hayashi (fully convective)
region. The first stage of migration
comes next, representing a time during which the planet
has roughly constant radius but increasing
effective temperature as the planet moves
closer to its star. The second stage of 
planetary migration, 
when the planet is
affected by torques from the star, is also
part of the increasing effective temperature
(i.e., inward) progression in the HR diagram. The third
stage of migration, due to mass loss, is outward
motion, seen in an HR diagram as decreasing
effective temperature.
During mass loss,
the planet's radius is increasing. The fourth stage of
evolution follows the cessation of mass loss,
during which time the planet again cools
and its luminosity decreases while at moving
to lower
effective temperatures (larger heliocentric 
distances) due to the remaining
tidal (star) torques.  Lastly,
all the external torques vanish,
and the planet continues to contract
and become less luminous, at constant $T_{eff}$.

\placefigure{hr}

Figure \ref{all_ma} compares
the final masses and heliocentric distances of our
migrating planets,
for various parameter combinations, with those
of the 
recently detected observations of extra-solar
planets.
This is not meant to be a complete
parameter study, but rather 
focuses in particular on cases interesting
to extrasolar planets and to
Jupiter. There are no Class I planets
in this figure, since they all lose all their
mass onto the star and do not survive.  Class II
planets (open circles) are
shown, and are similar to 51~Peg~b
and
$\upsilon$~And~b.
Class III, planets which 
do not lose any mass during their
evolution, comprises
the remaining planets, and reproduces
$\tau$~Boo~b, 55~Cnc~b, $\rho$~CrB~b,
47~UMa~b, and Jupiter. These
planets may
not have migrated far, and Jupiter
has probably moved radially
$<$ 1\,AU from
its formation location.
Not shown
are the extrasolar planets with high
eccentricities (16~Cyg~Bb, 70~Vir~b,
HD~114762~b), as our model does not 
produce planets in highly
eccentric orbits. In these systems,
it is likely that some post-migration
dynamical processing
has taken place\markcite{cochran}\markcite{linida}
(e.g., Cochran et al. 1997;
Lin \& Ida 1997).
The slight positive slope at small
heliocentric distances is suggestive
of an ``inner limit,'' and is caused
by the outward tidal torque, which is
active after the other torques have
vanished. There
are three parallel trends with
this slight positive slope, corresponding
to the three different $Q_{\star}$
values used in our study.
The magnitude of the tidal
torque is proportional to $M_{p}$
(eq.~\ref{dadt_star}), producing
the positive slope. Systems in 
which the tidal torque is active
long after the other torques, and which
have $Q_{\star}$ within the range
considered in this work, should not
have planets inward of this ``inner limit,''
and this late tidal evolution has the effect
of erasing any locational signatures
of previous evolutionary stages.
The preponderance of models which have 
final masses of 1\,$M_J$ reflects an 
excess of model runs with this initial mass,
in order to study Jupiter's evolution.
The curving trends of planets at $>5\,M_{J}$
and $>$ 1\,AU represent planets which are
massive enough to clear their disks before
they can migrate inward very far.
More massive planets clear their disks 
more quickly, so that the most massive
planet in our models, 13\,$M_{J}$, has a final
heliocentric distance which is only a few
percent different than its starting position.
In general, more massive planets do not 
move as far, although we have several examples
of large ($>$5\,$M_J$) planets at small heliocentric
distances, due to migrations is disks with
varying diskmasses or viscosities.

\placefigure{all_ma}

\section{Discussion \label{discussion}}

We find that a model of migrating planets
can explain the presence and locations of
not only the close companions
(at $\lesssim$ 0.1 AU), but also reasonably
reproduces other observed planets,
including Jupiter.  In the cases of 
a Class
II planets, the ultimate planet may be
only some 15\% of its initial mass. We
further find that the observed extrasolar
planets represent a subset of possible
outcomes of migrating planets.
Because of the steepness of the 
slope of $M_{f}/M_{i}$ in the 
Class II regime (see Figure~\ref{mfmi}),
very small changes in initial mass
in the Class II mass range result 
in large differences in final mass.
Moreover, as can be seen
in Figure~\ref{all_ma}, there is
a tendency for planets from both
Classes II and III to end up at
small heliocentric distances 
(0.03 to 0.05\,AU). This result
is caused by two separate physical
mechanisms: Class II planets are
moving away from the central star
as they lose mass. Class III planets
are still moving inward at the time
when the disk vanishes.
Therefore, the range of initial
masses which have small final
heliocentric distances is broader
than the critical mass range,
as some planets from Class III,
as well as all planets from Class
II, can end up at small heliocentric
distances. The result is a ``piling up'' of final
planets at small heliocentric 
distances: planets with a fairly
wide range of initial masses can
all end up at small orbital
separations.
The presence of several detected planets at 
small masses and small heliocentric 
distances may imply geneses in
disks with similar
physical characteristics;
or it may be the result of preferentially
ending up at small heliocentric distances,
despite somewhat different initial
conditions.
We cannot distinguish between these 
possibilities with the current model.

However, from Figure~\ref{all_ma}, we observe
that there seems to be a cutoff, in 
both heliocentric distance and final mass,
between Class II and Class III planets.
There are no
Class II planets, in $M_f - a_f$
space,
outside the boundaries defined by
final masses $\gtrsim 2 M_J$
and final separations $\gtrsim$ 0.05\,AU;
and there are no Class III planets within
that range.
This
segregation is
found over a wide range of disk parameters.
Therefore, we suggest that observed planets within
this region likely are Class II planets,
having lost some but not all of their
mass.
Certainly, with exactly the
correct disk parameters, Class III
planets could exist in this region.
However, for our fairly wide coverage
of parameter space, we do not find {\it any}
Class III planets at small heliocentric
distances and small masses, similar to
51 Peg b and $\upsilon$ And b. Therefore,
we believe that our model, with the
three important torques included, is relevant
and important in understanding the evolution
of some of the extra-solar planets, in that
these two planets fall within the region where
mass loss almost certainly is important.
Future observations of extrasolar
planets may well allow us to constrain
some disk and star parameters.

Our model spans a large
part of multi-dimensional parameter
space, as we have five major physical
parameters which can affect the fates
of migrating planets. 
We have examined the statistical
results for various slices through
our five-dimensional parameter space.
Regardless of which slice (comparing
diskmasses, viscosities, etc.) is chosen, for
all samples of reasonable size, the 
statistical results are roughly
equivalent, another measure of
the importance and validity of
our model. We find that, for
planets with initial integral
masses between 1 and 5 $M_J$,
approximately half of the planets
lose all their mass onto the central star (Class
I planets),
and approximately half survive (Class II or III).
Typically,
a few percent of planets end up as
Class II planets, which have lost some 
mass.
The fact that half the initial planets
lose all their mass may imply that half the
normal-type, giant planet forming systems
have lost planets onto their central
stars. Of the planets that survive, 
between one quarter and one third typically
have heliocentric distances greater than
1\,AU. Therefore, while Jupiter is consequently
not a special case, it is also
not the most probable outcome.

The percent of 
planets (Class II or III) which end up
at small heliocentric distances ($\lesssim$0.1 AU)
is 3 - 5\%. We find it intriguing
that this rate of producing companions at
small heliocentric distances is similar to
the rate of discovering planets at small
heliocentric distances\markcite{coolstars}
(Marcy \& Butler 1997). Our model predicts 
that one quarter to one third of giant
planet forming systems should have Jupiters
at $>$ 1\,AU; this percentage is much higher
than the observational success rate for planets
in large
heliocentric orbits. However, Jupiter-mass
planets in large ($\sim$5\,AU) orbits are
the most difficult to detect with radial velocity
searches, and in many cases are
beyond detection. However, in the near future,
better techniques on bigger telescopes
should provide improved data sets in this range,
and higher observational success rates can be
expected.
 
Planets in our theoretical model fall into one
of three classes, depending their mass loss. However, 
from the standpoint of observers and observables,
any Class III planet at a small heliocentric 
distance would be indistinguishable (at this time)
from a Class
II planet at the same distance and mass. Although
these two planets have had very different evolutions,
their observable characteristics (e.g.,
mass and heliocentric distance)
would
be quite similar.
Therefore, in terms of observations,
the most sensible classes for extra-solar
giant planets would be to
group planets which are close to their stars,
regardless of mass loss history. In this
``observer's scheme,'' the planet's evolution
is ignored, but the definitions are
more sensible for today's detection technologies. 
However, in terms of origins of planetary
systems,
classifying planets according to their
evolutions, 
as we have done, is more sensible.

In our model, planets are considered
to have disappeared completely once
their masses become less than 0.4 $M_J$.
In reality, of course, mass loss 
continues below this cutoff, and a 
stage could be reached in which all
of the gaseous envelope of a giant planet
was stripped away, leaving a rocky
core. Jupiter's core is thought to
be $\lesssim~0.1~M_J$, or $\lesssim$~30~Earth masses
(1~Earth mass~=~1~$M_\oplus$~=~$6\times10^{27}$~g),
and made of dense,
rocky material (Hubbard 1984).  If such
a remnant body were uncovered by
atmospheric mass loss onto the central
star, the remaining core, whose
density is much higher than Jupiter's
density overall, would not be in a 
mass overflow regime, and could continue
to migrate or survive at this mass. In
other words, a temporary halting of mass
loss would exist when the core is 
uncovered, and therefore there might be
a plateau in final masses at around the
mass of Jupiter's core. Thus, the final
mass and semi-major axis distribution of 
extrasolar planets may include a family
of planets with masses approximately equal
to Jupiter's core, at small distances;
these bodies would be rocky, as opposed to
primarily gaseous. Detection of
30 $M_\oplus$ bodies at small distances
might therefore provide evidence
for Roche lobe overflow
from primarily gaseous giant
planets, representing remnant cores
of the larger bodies. These smaller mass planets would
not necessarily have been detected, given
current radial velocity sensitivies, depending on
heliocentric distance (Marcy et al. 1997).

How important are the three torques 
which we consider? In Figure~\ref{nono},
similar to Figure~\ref{at},
we show the results of calculating
a planet's migration in the nominal disk
with two of the three torques which we
include in our model.
We can consider a system in which mass
loss is not present, e.g., the planets
are point-mass planets, simulating
rocky planets, as compared
to the nominal gaseous Jupiter used
in our standard model.
For systems in which mass loss is not present
but stellar tidal torques are,
planets can approach the star
much more than in the nominal case.
Ultimately,
stellar tidal torques are not enough to reverse
a planet's motion, so that unless the disk 
dissipates, these planets will dissappear 
onto their central stars. 
Rocky planets (i.e., planets
without mass loss) are an interesting special
case, but probably do not represent
a physically plausible creation
mechanism for giant planets.
In a system in 
which stellar tidal torques do not exist,
a planet loses mass on a timescale equivalent
to its migration timescale. and thus very 
few planets are rescued by the disk's evaporation.
Planets in systems without tidal torques
would only rarely survive once reaching 
small heliocentric distances.
However, as is demonstrated
in Figure~\ref{nono}, the combination of all
three torques extends a planet's lifetime in 
proximity to its central star such that 
a reasonable fraction of planets can survive
at small heliocentric distances. With only
two of the three torques acting, the already
narrow critical mass range becomes vanishingly
small.

\placefigure{nono}

The effects of several key parameters
on the fates of planets can be seen in
Figure~\ref{trends}, which 
plots the boundary (in initial
masses) between Class II and Class III
planets, versus varied model parameter.
This boundary is effectively
a proxy for the
initial mass for which planets
will survive at small heliocentric
distances, at $M_f \leq M_i$.
Larger viscosities and larger
diskmasses cause the critical mass range to
shift to higher masses by creating larger
disk torques on the migrating
planets, so that planets migrate faster.
The result
is that planets at small heliocentric distances
could have larger masses than is the case
for our nominal model. Conversely, larger
initial heliocentric distances allow for planets
with smaller masses to reach small orbital
separations, since the time to
migrate to small heliocentric distances
is longer, and fewer planets have enough
time to migrate in close and lose all their
mass before the disk dissipates. Larger values of $Q_\star$
also move the critical mass range to larger
initial masses, because the tidal torque
is lessened, so planets feel less strong
outward torque due to tides. The dependence
on $Q_\star$ is fairly weak, however.
Lastly, the relevant
timescales cause their obvious changes in
the critical mass central value: smaller
disk lifetimes move the critical mass
range to smaller masses, and larger
disk lifetimes cause the critical mass range
to move to larger initial masses.
Shorter effective disk lifetimes
can allow planets which fall into Class I in
the nominal disk to fall into either Class II
or III in a disk with a shorter
lifetime, depending on how the migration time
compares to the shorter disk lifetime.

\placefigure{trends}

The effect
of requiring larger initial masses to
survive is twofold. Firstly, it means that 
fewer objects
will survive if larger initial masses are
needed to survive. This is simply because the
fraction of planets with initial masses below
this cutoff is larger, so that fewer of the
total number of planets survive until $10^{10}$\,years.
In other words, the mass range which bounds
Class I is larger, so more planets fall into
Class I and do not survive.
Secondly, a wider range of final masses
at small heliocentric distances
is possible.
Since there is a piling up of planets at small 
heliocentric distance at about the initial mass
for which $M_f=M_i$ (i.e., the boundary between
Classes II and III), the effect of moving this
boundary to larger initial masses is to create
a piling up at larger final masses. Furthermore,
planets at small heliocentric distances can
have final masses less than their initial
masses; so that if the mass for which $M_f=M_i$
is larger than nominal, then the range of masses
between $M_f=M_i$ and $M_f=0.4\,M_J$ is larger,
and planets at small heliocentric
distances can have a wider range of final
masses.
For example, in the nominal
disk, it is not possible to have a 5~$M_J$ planet
reside at small heliocentric distances. However,
at larger $\alpha$'s, a 5 $M_J$ planet
can end up at $a_f\ll$ 1\,AU. Therefore,
a wider range of fates is possible when
the cutoff mass is larger; however, fewer
total planets survive when the cutoff mass is larger.

Future elaborations in our model include
examining non-conservative 
forces
which may control the width of the
critical mass range (for example,
stellar wind and magnetic field
interactions).
In our model,
we have taken the effect of
atmospheric evaporation of 
the planet to be small.\markcite{guillot}
Guillot et al. (1996)
find that, for 51~Peg~b, 
in its close orbit, thermal evaporation
(classical Jeans escape) is insignificant,
and that nonthermal evaporation could
cause a total mass loss, over the lifetime
of the system, of less than 1\% of the
mass of the planet. In our model, planets
can migrate to locations 
inside of 51 Peg b's heliocentric distance,
thus having higher temperatures and
increased stellar fluxes.
Therefore, while the net effect is
probably small relative to mass loss
and tidal torques, further refinement of
the model necessitates re-examination
of stellar wind effects.

Our Solar System, of course, should not
have been immune to giant planet migration.
If Jupiter has moved somewhat from its initial
location, there are possible ramifications for
other bodies in our Solar System.\markcite{malhotra}
Liou \& Malhotra
(1997) have explored the dynamical
effects on asteroids of a migrating Jupiter, and show 
that a depleted region in the outer
asteroid belt can be explained by 
Jupiter migrating inward 0.2\,AU.
One model run produces a 1\,$M_{J}$
planet which forms at 5.2\,AU in a disk
of less than nominal mass and
$\tau_{disk}$ of $10^{6}$\,years,
and halts its inward migration
at 4.48\,AU.
This disk has a shorter than nominal
lifetime (that is, time after the 
planet has fully formed), and a less
than nominal diskmass.
These parameters
may represent a disk which is 
already dissipating as the planet starts
to migrate, implying a formation 
time for Jupiter of nearly $10^7$\,years, if
our circumstellar disk lasted the nominal
$10^7$\,years after protostar collapse.
By estimating Jupiter's
migration distance, we may be
able to start describing
the circumstellar disk conditions
in our early Solar System.

It has 
been proposed that orbital resonances 
with Jupiter may play an important
role in the formation of asteroids
and terrestrial planets\markcite{wetherill91}\markcite{wetherill92}
(Wetherill
1991, 1992). Jupiters
which migrate through
several AUs must certainly
disrupt the formation of 
smaller planetary bodies
as orbital resonances sweep through
the circumstellar disk,
and it
is almost certain that any Earth-like
planets which were forming at 1\,AU
in the 51 Peg system during the time that 51 Peg b
migrated inward were
disrupted. The gap in the disk fills
in on the viscous time scale after
the passage of the planet;
if the disk fills in completely enough
and rapidly enough after giant planet 
passage, it is possible that terrestrial-sized
planets could form in the wake of planetary migration,
although the orbital resonances from 
a very close companion would be very
different and probably much less
important than from a Jupiter-mass planet
at 5.2\,AU. 

\section{Conclusions \label{conclusions}}

Planets in circumstellar
disks are subject to various
torques which can cause the planets
to migrate inward.
We present results from a 
self-consistent quantitative model
which describes evolution and
migration of giant planets.
Some planets
migrate very rapidly
and disappear due to mass 
loss in less time than the
disk lifetime. Other planets
or planets under other disk 
conditions can be left in
various orbits, from very close to
the star to very close to where they
started, and with masses equal to
or less than their initial masses.
There is a wide range of possible 
fates (heliocentric distance
and final mass) for migrating Jupiters.
A migrating planet has one of three
possible fates:
losing mass
to the point of extinction 
(Class I); surviving
but having lost some of
its mass onto the central star
(Class II); or
surviving after having
migrated and not lost any 
mass (Class III).
Some planets in this last
class may have migrated only
a small radial distance.
The width of Class II is a function
of a variety of disk and star
parameters. The region
of $M_f - a_f$ space
delimited by 
$M \lesssim$ 2\,$M_J$ and
$a \lesssim$ 0.05\,AU is
populated {\it only} by
Class II planets, and
includes 51 Peg b and $\upsilon$
And b.
All observed extrasolar close
companions fall within the 
population of surviving planets
in our model (Classes II and III),
suggesting that
this mechanism may explain the
detections of planets very close
to their stars. We also
suggest a ``piling up'' effect
at small heliocentric distances,
and that planets may
survive close to their stars
in numbers disproportionate to
initial distributions. 

An observational way of assessing the significance of mass loss for
the orbital evolution of giant planets would be to look at the orbital
distances of planets around different types of stars. If there is no
systematic effect (for example, due to different chemical compositions), the
internal evolution of the planets should be the same around any star.
But around more massive stars (e.g. F type stars), mass loss would occur at
larger distances, whereas less massive ones (e.g. K type stars) should be
able to retain some planets at smaller heliocentric distances. 
Furthermore, since all planets in our study with
small ($< 2\,M_{J}$) mass and small (0.05\,AU)
orbital distances have lost mass, another
possible test of our mechanism in
the future is to study the
compositions of the planetary
remnants which survive and their 
stars to detect
chemical signatures of mass transfer.
If signatures exist, observationally defining the 
boundary between Class II and Class III 
planets would be possible, thereby testing
the model and, in turn, constraining the
initial conditions of the circumstellar
disk in which the planet formed.

\acknowledgments

The authors wish to thank
Didier Saumon, Chris Chyba,
and Doug Lin
for stimulating discussions.
We also acknowledge useful suggestions
from an anonymous referee.
D. T. is supported by an NSF 
Graduate Research Fellowship.
T. G. is supported by
a Training and Mobility of Researchers 
grant from the European Community.
This work is supported in part
by the NASA Origins of Solar Systems
Program, grant NAG5-4051.

\clearpage

\begin{deluxetable}{ccccc}
\small
\tablecaption{Properties of Observed Extrasolar Planets \label{table1}}
\tablehead{
\colhead{Star} & \colhead{a\,(AU)} &
\colhead{$M \sin i\,(M_{J})$} & 
\colhead{e} & \colhead{Ref.}}
\startdata
51 Peg & 0.05 & 0.46 & $<$ 0.01 & 1,2 \nl
$\upsilon$ And & 0.057 & 0.68 & 0.1 & 3 \nl
55 Cnc & 0.11 & 0.84 & 0.05 & 3 \nl
$\rho$ CrB & 0.23 & 1.1 & 0.03 & 4 \nl
16 Cyg B & 1.6 & 1.5 & 0.63 & 5 \nl
47 UMa & 2.1 & 2.39 & 0.03 & 6 \nl
$\tau$ Boo & 0.046 & 3.87 & 0.02 & 3 \nl
70 Vir & 0.43 & 6.6 & 0.4 & 7 \nl
HD 114762 & 0.41 & 10 & 0.33 & 8 \nl
\enddata
\tablecomments{
The mass of planetary companions, as 
determined by radial velocity searches,
can only be determined to within a 
factor of $\sin i$, where $i$ is the
inclination of the planet's orbit
with respect to the observer.}
\tablerefs{
(1) Mayor \& Queloz 1995
(2) Marcy et al. 1997
(3) Butler et al. 1997
(4) Noyes et al. 1997
(5) Cochran et al. 1997
(6) Butler \& Marcy 1996
(7) Marcy \& Butler 1996
(8) Latham et al. 1989}
\end{deluxetable}

\clearpage

\begin{deluxetable}{ccccc}
\scriptsize
\tablecaption{Model Results: Various Initial Masses in the Nominal Disk
	\label{nominalmass}}
\tablehead{
\colhead{$M_{i}\,(M_{J})$} &
\colhead{$M_{f}$\,($M_{J}$)} &
\colhead{$a_{f}$\,(AU)} &
\colhead{$t_{stop}$\,(y)} &
\colhead{Class}}
\startdata
1.00&0.39&0.033&$1.75\times10^{6}$&I \nl
2.00&0.50&0.030&$2.42\times10^{6}$&I \nl
3.00&0.39&0.047&$5.67\times10^{6}$&I \nl
3.30&0.37&0.037&$8.88\times10^{6}$&I \nl
3.35&0.37&0.036&$9.73\times10^{6}$&I \nl
3.36&0.38&0.037&$5.97\times10^{6}$&I \nl \tableline
3.37&0.47&0.032&$10^{10}$&II \nl
3.38&0.63&0.031&$10^{10}$&II \nl
3.39&1.04&0.033&$10^{10}$&II \nl
3.40&1.86&0.037&$10^{10}$&II \nl \tableline
3.41&3.41&0.040&$10^{10}$&III \nl
3.42&3.42&0.040&$10^{10}$&III \nl
3.43&3.43&0.040&$10^{10}$&III \nl
3.44&3.44&0.040&$10^{10}$&III \nl
3.45&3.45&0.040&$10^{10}$&III \nl
3.50&3.50&0.040&$10^{10}$&III \nl
3.60&3.60&0.041&$10^{10}$&III \nl
3.80&3.80&0.041&$10^{10}$&III \nl
4.00&4.00&0.45&$10^{10}$&III \nl
4.50&4.50&1.63&$10^{10}$&III \nl
5.00&5.00&2.22&$10^{10}$&III \nl
6.00&6.00&2.89&$10^{10}$&III \nl
7.00&7.00&3.30&$10^{10}$&III \nl
8.00&8.00&3.57&$10^{10}$&III \nl
9.00&9.00&3.78&$10^{10}$&III \nl
10.00&10.00&3.98&$10^{10}$&III \nl
\enddata
\tablecomments{
The nominal disk has the 
following parameters: $\alpha=5\times10^{-3}$,
$Q=1.5\times10^{5}$,
$\tau_{disk}=10^{7}$\,years,
$\tau_{spin}=10^{8}$\,years,
$M_{disk}=1.1\times10^{-2}\,M_{\sun}$,
and initial distance = 5.2\,AU.
The subscripts ``i'' and ``f''
refer to initial and final values,
respectively. The final values
are the values at $t_{stop}$,
which is the time when the 
model stops: either at $10^{10}$\,years,
or when
the planet's mass is less
0.4\,$M_{J}$, which is the lower mass
cutoff for planets
in our model, or else when
the model stops converging
(usually as the planet's mass
nears 0.4\,$M_{J}$).}
\end{deluxetable}

\clearpage

\clearpage




\figcaption{Gap size versus initial planetary mass,
for two different viscosities and two
different models. Planets orbiting
in disks open gaps. The size of the gap depends
on the mass ratio $M_p/M_{disk}$ and also on the
viscosity of the disk. This figure compares our 
results (circles) using the local dissipation approximation
of Lin \& Papaloizou (1986) to the results
of Takeuchi et al. (1996), who use a more computationally
intense WKB approximation (squares). Open symbols represent 
our nominal viscosity of $\alpha=5\times10^{-3}$, and filled
symbols represent a high viscosity, with $\alpha=10^{-2}$.
Gap sizes are plotted
for t $\approx 10^4$\,years.
The gap size results produced by
the two models are similar, so
that using the simpler local dissipation
approximation in our calculations is appropriate.
\label{gap}}

\figcaption{Heliocentric distance versus time,
for various initial masses, in the nominal disk.
Evident here are all three classes: Class I,
which lose all their mass ($\leq3.36\,M_{J}$),
Class II, with initial masses
$3.36\,M_{J}<M_{p}<3.41\,M_{J}$, and Class III,
with initial masses $\geq 3.41\,M_{J}$.
See text for discussion of these
three classes. The outward
motion of the surviving planets
between $10^{7}$ and $10^{8}$\,years
is because the (outward) tidal
torque is the only torque acting
on the planets during this time.
The shaded region in the lower left
represents the radial extent of the
central star.
\label{at}}

\figcaption{Planetary mass versus time, for
various initial masses, in the nominal disk.
Similar to Figure~\protect\ref{at}. 
The three classes of planets can clearly
be seen.
\label{mt}}

\figcaption{Planetary mass versus heliocentric
distance, for various initial masses
in the nominal disk. Similar to
Figures~\protect\ref{at} and \protect\ref{mt}.
Some planets do not lose any mass during
their migrations. For those planets that
do, the planetary radius increases with
decreasing mass. Since a planet which is
losing mass stays at a distance at which its
Roche radius is equal to its planetary 
radius, planets losing mass move away from
the central star. Thus, there is
a negative slope in $M-a$ space;
the slope of this line has the same
sign as $\alpha_{p}$.
Late (t $> 10^8$\,years) outward motion at 
a constant mass of 1.86\,$M_{J}$
by the planet
which initially has mass 3.4\,$M_{J}$
is caused by tidal torques between
$10^{7}$ and $10^{8}$\,years.
\label{ma}}

\figcaption{Final mass versus initial
mass, for model planets in the nominal circumstellar
disk. The vertical lines at $M_{i}$ = 3.36
and 3.41\,$M_{J}$ represent the boundaries
between Classes I and II, and II and III,
respectively, and define the critical mass
range as the area between the lines.
All Class I planets lose all
their mass before the disk dissipates, and
hence have a final mass of 0. Planets
in Classes II and III survive the
migration and mass loss process. Planets
in Class III do not lose any mass (hence 
the line has slope 1 for $M_{i}>3.41\,M_{J}$).
\label{mfmi}}

\figcaption{HR diagram comparing the
evolution tracks of Jupiter (grey line), of a 1\,$M_{J}$
planet at 0.05\,AU from its sun (dashed
line, from Guillot et al. 1996),
and of a planet of initially 3.39\,$M_{J}$ and 5.2\,AU
(black line).
The migrating planet has a final mass of 1.04\,$M_{J}$.
Numbers next to dots 
represent the log of time, in years, of
the planet's evolution. All evolution tracks start at high
luminosity and small effective temperature (large radius). 
The period of increase of the effective temperature at almost
constant radius corresponds to the inward migration
stage. The evolution of the planet is
then dominated by the increasing stellar heating.
The subsequent decrease of effective temperature
with roughly constant luminosity is due to mass
loss and consequent increase of the orbital radius.
The final stage of the migrating planet's 
life is cooling at constant temperature, here
shown as the vertical line downward 
from $10^{8}$\,years. The
hatched and cross-hatched regions indicate
the Hayashi-forbidden regions for
3 and 1\,$M_{J}$ planet, respectively. Lines
of constant planetary radii are shown, in units
of $R_{J}$ ($7\times10^{9}$\,cm). The luminosity
of Jupiter is $10^{-9}\,L_{\sun}$, or a few times
$10^{24}$\,ergs/sec. The diamond in the lower
right corner shows Jupiter's present
characteristics.  
\label{hr}}
 
\figcaption{Final mass versus heliocentric distance,
for model runs plus observed extrasolar planets.
This figure includes results from many
model runs with different
physical parameters varied,
but is not meant to be an evenly gridded or complete
parameter study. See Figure~\protect\ref{trends}
for ranges of parameters used. Initial
masses are 0.5\,$M_J\leq M_p\leq 13\,M_J$. 
Open circles show planets which lose 
some mass during their 
migration, but survive (Class II).
Filled circles indicate models in which
the planet does not lose any mass during its evolution
and migration (Class III). Triangles show observed
extrasolar planets. We find a wide range
of final heliocentric distances and masses 
for migrating planets.
Observed planets included are the following: 51 Peg
b, $\upsilon$ And b,
55 Cnc b, $\rho$ CrB b,
47 UMa b, and
$\tau$ Boo b. See Table~\protect\ref{table1}
for properties of the observed extrasolar planets.
Jupiter is the diamond at 5.2 AU and 1 $M_{J}$.
\label{all_ma}}

\figcaption{Figure demonstrating the 
importance of the three torques we consider.
Here we plot heliocentric distance
versus time, as in Figure~\protect\ref{at},
but with expanded axes.
Shown are model results for planets with
initial masses 3.4 and 3.5\,$M_J$, for
scenarios with and without stellar tidal
torques. Also shown is a point mass [rocky] 1\,$M_J$ planet
in a system which does have tidal torques.
Tidal torques prolong, by up to 
a factor of ten, the length of time that
planets can stay at small heliocentric 
distances without losing all their mass. Thus, a planet
with initial mass of 3.4\,$M_J$, without
tidal torques, loses all of its mass quickly,
and is a Class I planet. With tidal
torques in the system, this planet survives 
as a Class II planet. 
A planet with initial mass 3.5\,$M_J$ is a Class
III planet in the nominal disk and a Class I
planet in a system without stellar tidal torques.
The point mass planet,
an analog for a rocky planet, is included
to show the effects of having stellar
tides present, but not mass loss. In this
case, a 1\,$M_J$ planet is a Class III planet,
whereas in the nominal case of a gaseous giant
planet, a 1\,$M_J$ planet loses all of its mass
by $2\times10^6$\,years (see Figure~\protect\ref{at}
and Table~\protect\ref{nominalmass}).\label{nono}}

\figcaption{Boundary, in initial
masses (in $M_J$), between Class II and
Class III, over a range of varied parameters.
Increasing the parameters $Q_\star$, diskmass,
and viscosity cause this
boundary to move to larger initial masses.
Thus, in disks with higher viscosities (for
example), fewer planets would survive,
although a wider range of planets at
small heliocentric distances could
result (see text for discussion).
Increasing
the initial heliocentric distance of the planet
allows for smaller planets to survive with
no mass loss, as the 
inward migration time is longer.
Shorter disk lifetimes allow planets
with initially smaller masses to survive.
Nominal values
are the following:
$\alpha=5\times10^{-3}$;
$M_{disk}=1.1\times10^{-2}\,M_{\sun}$;
disk lifetime~$=10^{7}$\,years;
$Q_\star=1.5\times10^5$;
and $a_i=5.2$\,AU
(see Section~\protect\ref{disk} for discussion
of these parameters).
All of 
these physical parameters are varied
from their nominal values to model 
disks with different properties.
\label{trends}}

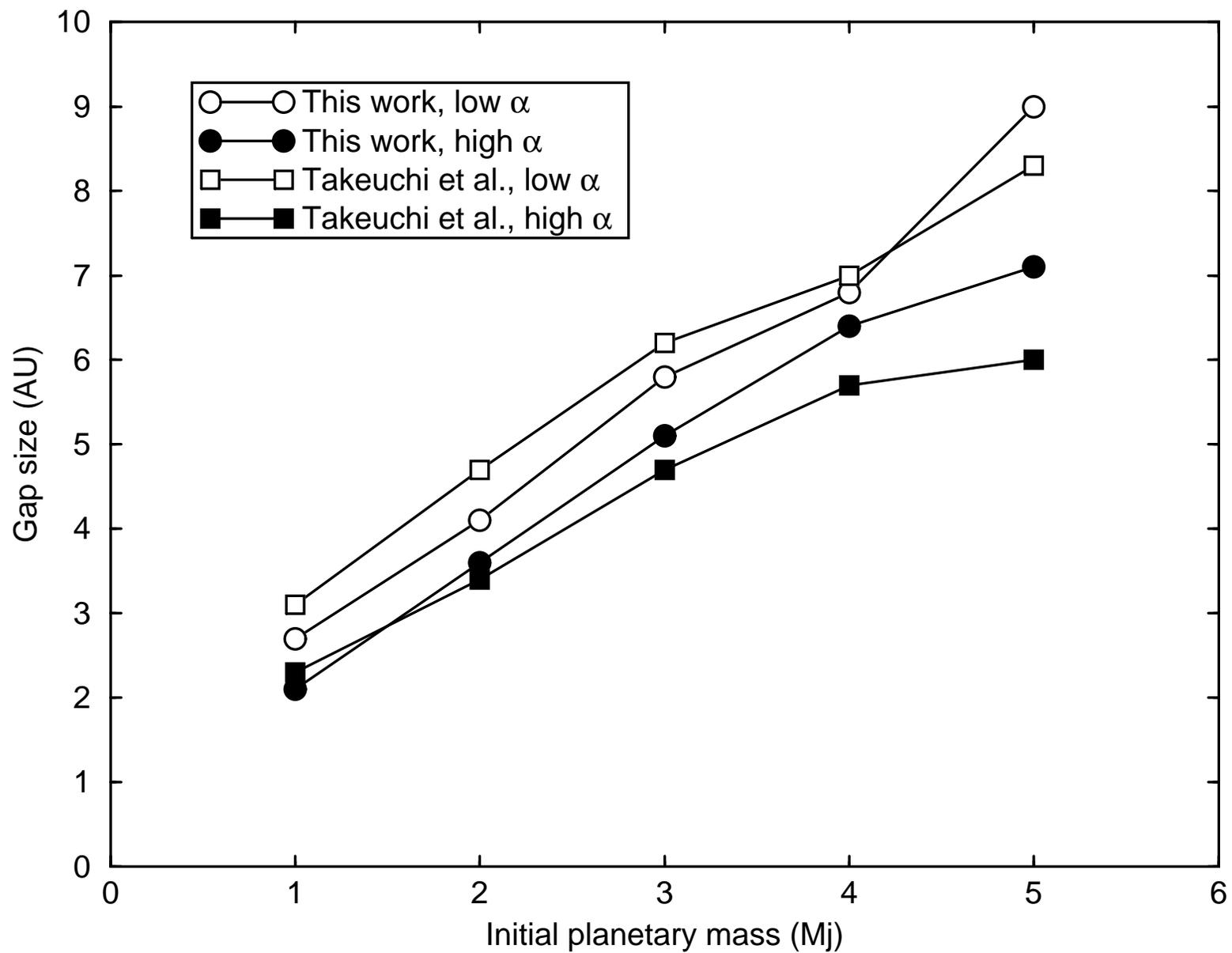
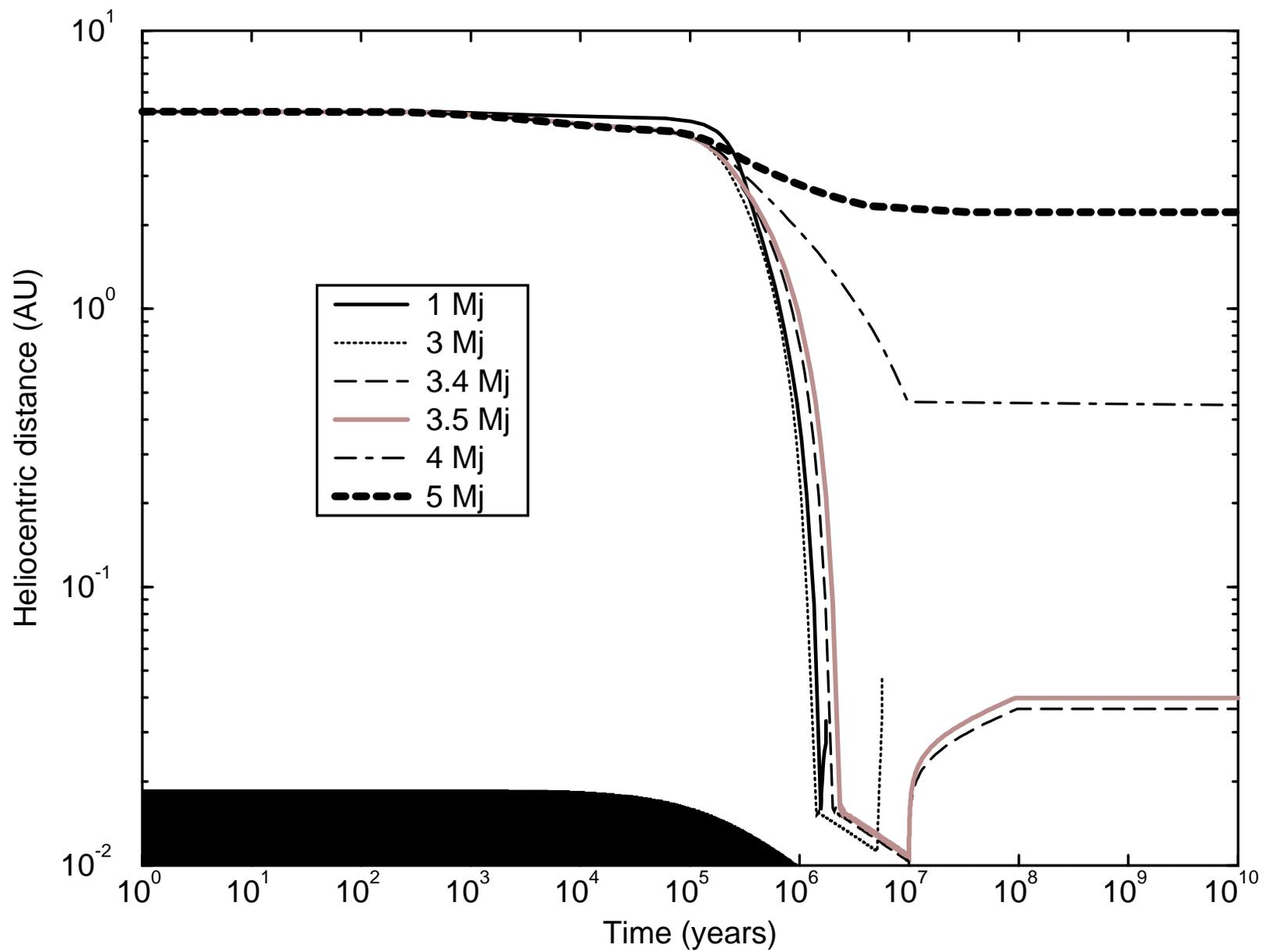
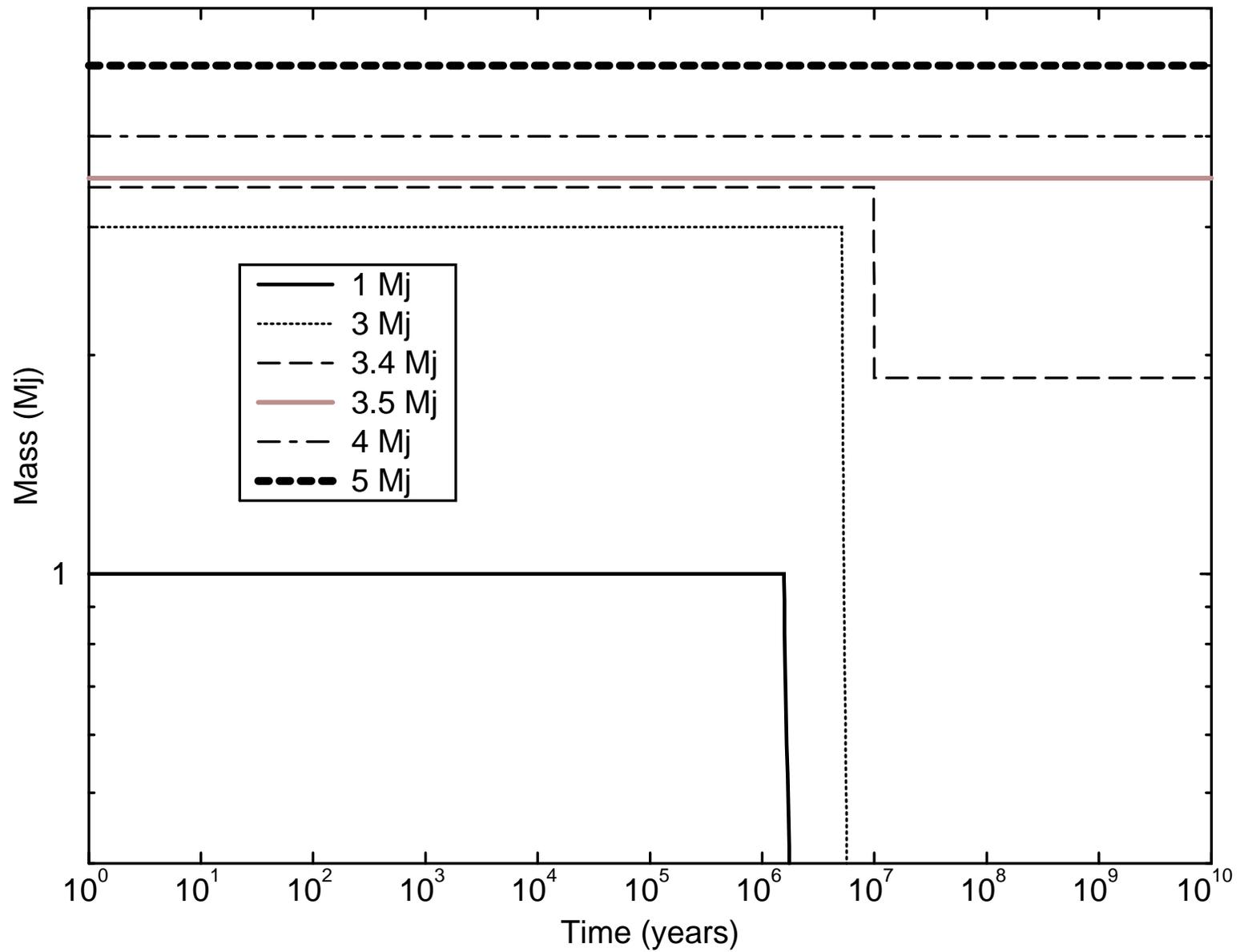
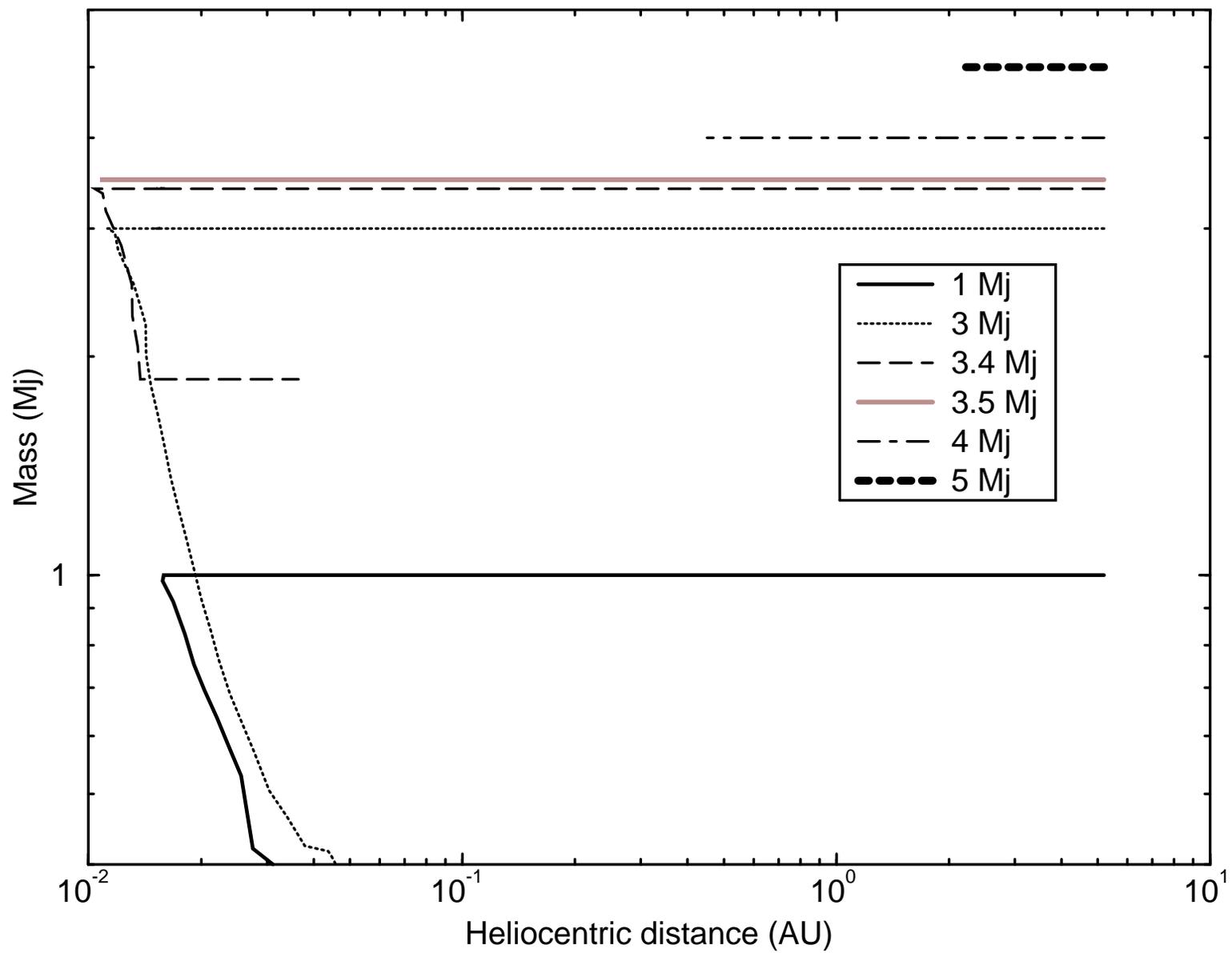
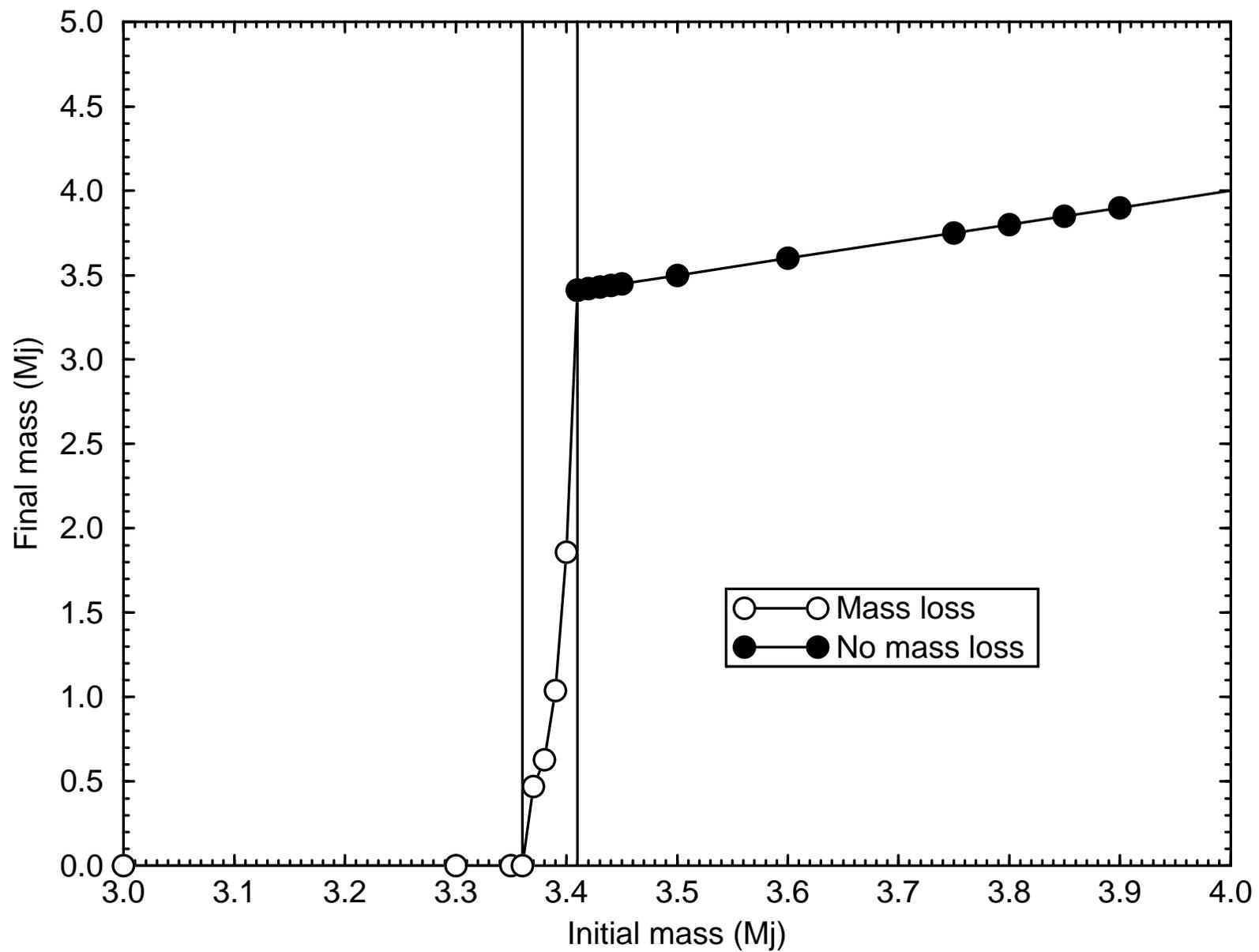
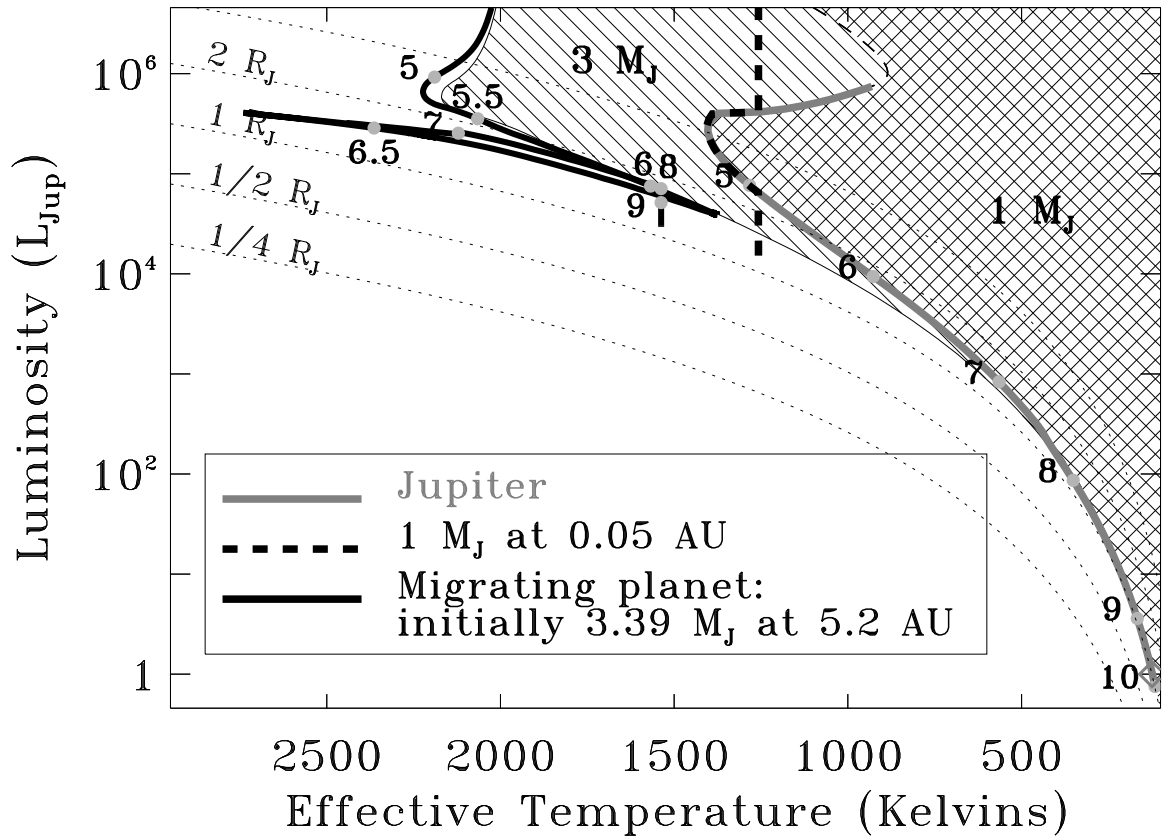
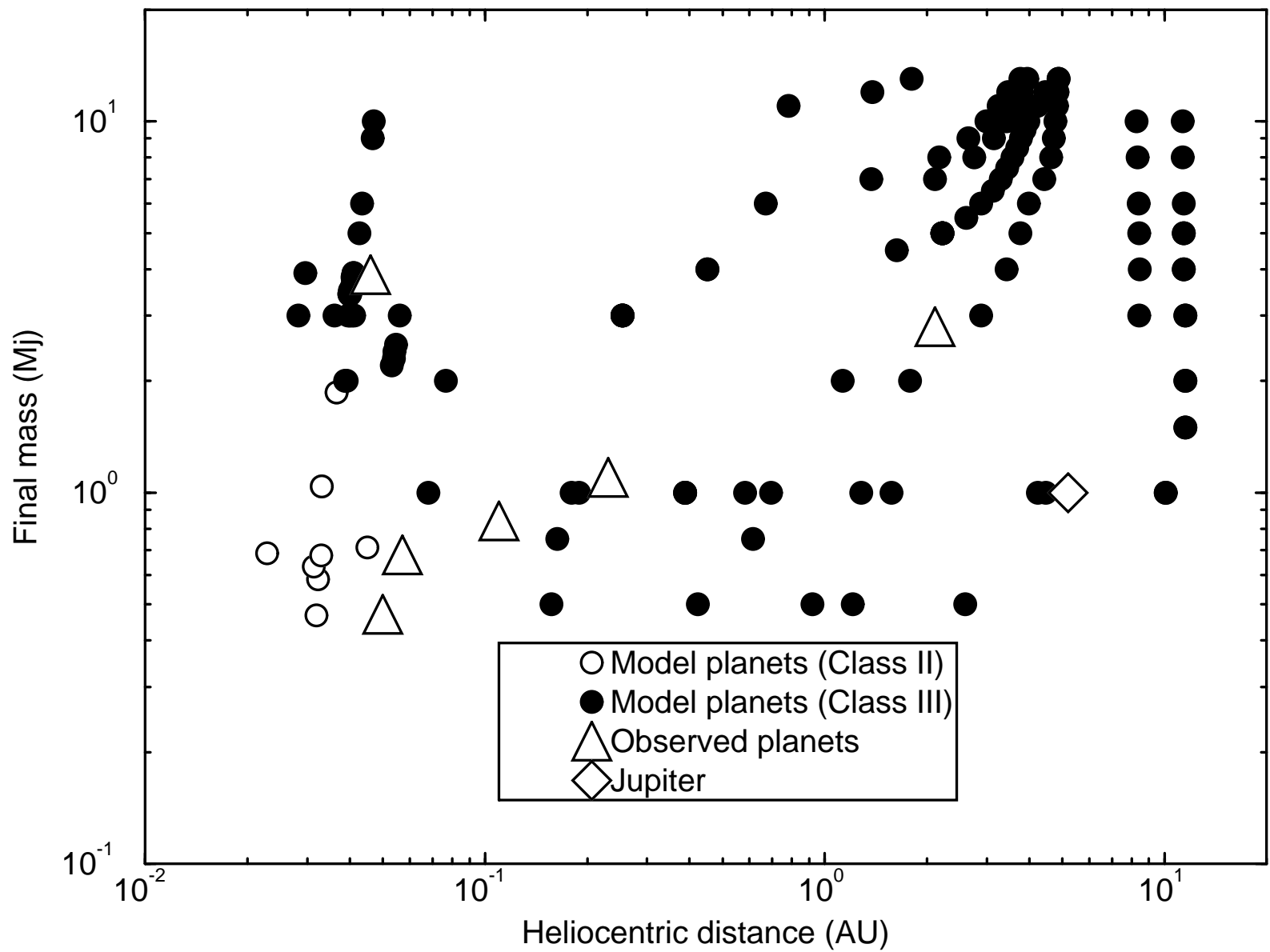
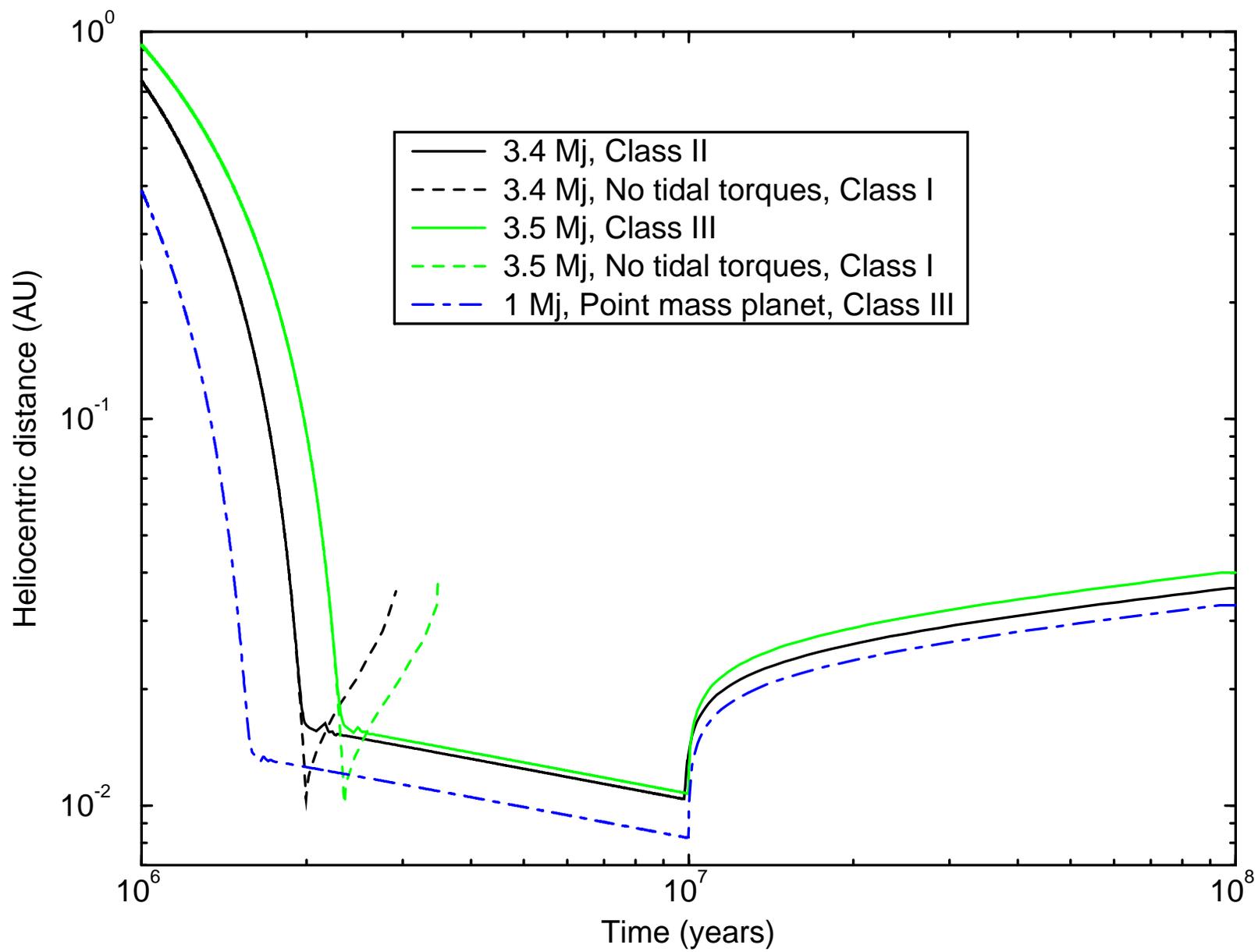
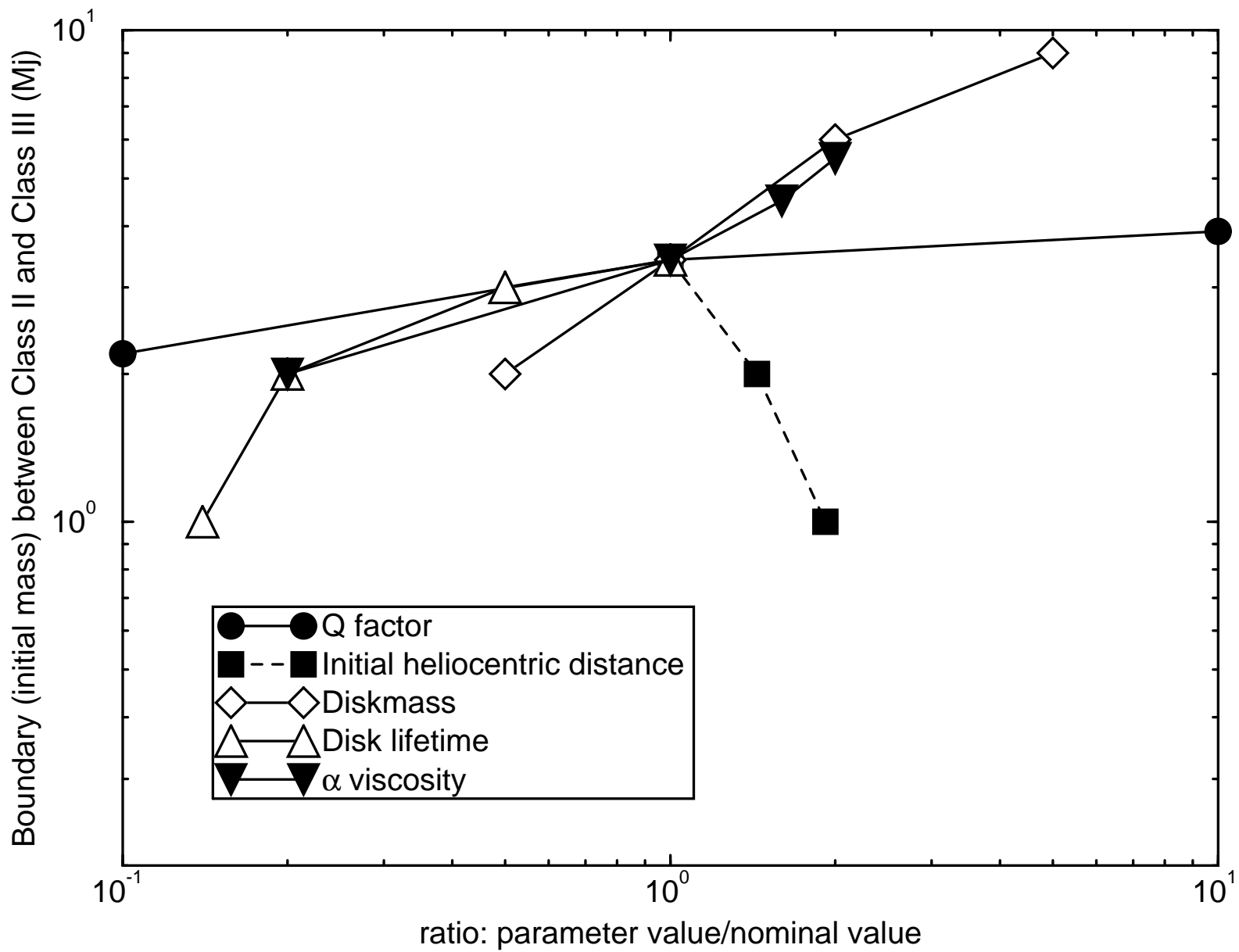
\end{document}